\documentstyle{aipproc}

\begin{document}
\title{Making glue in high energy nuclear collisions}

\author{Alex Krasnitz$^*$ and Raju Venugopalan$^{\dagger}$}
\address{$^*$UCEH, Campus de Gambelas, Universidade do Algarve, Faro, P-8000, 
Portugal\thanks{AK's work supported by Portuguese Funda\c{c}\~{a}o para 
Ci\^{e}nca e a Technologia, grants CERN/S/FAE/1111/96 and CERN/P/FAE/1177/97}\\
$^{\dagger}$Physics Department, BNL, Upton, NY 11973\thanks{Invited talk by RV
at the VIII Mexican School of Particles and Fields. RV's work supported by
DOE Nuclear Theory at BNL.}}

%\lefthead{LEFT head}
%\righthead{RIGHT head}
\maketitle

\begin{abstract}

We discuss a real time, non--perturbative computation of the
transverse dynamics of gluon fields at central rapidities in very high
energy nuclear collisions.

\end{abstract}

\section*{Introduction}
%\[
%\widehat{a} + \widehat{ab} + \widehat{abc} + \widehat{abcd}
%\]
%
%%\show\frak
% 
%\[
%%      {\bf x}^{\bf x} \triangleq z 
%      {\bf x}^{\bf x}\triangleq{z} \tensor{T} \frak{E^E}=\frak{mc}^2
%%      {\bf x}^{\bf x}\triangleq {z} \tensor{T} \frak{E}=\frak{mc}^2
%\]
% 
%\[
%{\Bbb {NQRZ}} \qquad \because \eth\ggg\bigstar \therefore\blacktriangleright\rightsquigarrow \blacksquare
%\]
% 
This year (1999) the Relativistic Heavy Ion Collider (RHIC) at
Brookhaven will begin colliding beams of gold ions at center of mass
energies of 200 GeV/nucleon.  In slightly over 5 years from now, the
LHC collider at CERN will do the same at center of mass energies of
about 5.5 TeV/nucleon. At these energies, the appropriate basis to
describe the colliding nuclei is that of partons, the quarks and
gluons that constitute a nucleon, rather than a hadronic basis. An
objective of the above mentioned experiments is to determine
whether the partons, confined in nucleons prior to the collision, are
liberated after the collision to form fleetingly, in the relatively
large nuclear volume, an equilibrated state of matter popularly known
as the quark gluon plasma.

Clearly, the space--time evolution, and possible equilibration, of matter
produced in a nuclear collision must depend on the initial
conditions. These are given by the parton distributions inside each of
the nuclei. In perturbative QCD, the ``factorized''
expression for the multiplicity or energy distribution of a high $p_t$
jet, with $p_t \approx \sqrt{s}$, is obtained by convolving the
parton distributions in each of the nuclei, at the hard scale of
interest, with the elementary parton--parton cross section. If $x\equiv 
p_t/\sqrt{s}$ is not too small, the factorized expression is reliable.    
With the measured nuclear structure functions, one can then compute the 
multiplicity and energy distributions of the jets produced~\cite{kajlandesk}.

However, for $x<<1$ (corresponding to the transverse momentum range
$\Lambda_{QCD}<<p_t<<\sqrt{s}$) the factorization formula for energy
and multiplicity distributions breaks down. Simply put, this is
because partons in one nucleus can resolve more than one parton in the
other~\cite{Rajcomment}. The parton densities in the nuclei become
very large and may even saturate at sufficiently small x. The regime
of high parton densities is a novel regime in QCD where, although the
coupling constant may be small, the fields strengths are large enough
for the physics to be non--perturbative~\cite{GLR}.

The precise x value at which the above mentioned leading 
twist factorization breaks down is not clear. There are hints from from HERA
that parton saturation is already seen in the data 
for $x\approx 10^{-4}$ and $Q^2 \approx 
4$ GeV$^2$~\cite{Mueller99}. If this result is robust, similar effects may 
be seen in nuclei at larger values of x, even $x \sim 10^{-2}$. 
Their relevance for RHIC (and especially LHC) cannot then be ignored.

The effects of high parton densities in the central rapidity region of
nuclear collisions can be studied in a model which is based on an
effective field theory (EFT) approach to QCD at small
x~\cite{Mobsters}. The model describes the time evolution of gauge
fields in a nuclear collision. It takes into account,
self-consistently, interference effects (which are also
responsible for shadowing in deeply inelastic scattering) that become
important at small x. Another nice feature is that it provides a
space--time picture of the nuclear collision. This feature would be
extremely useful if the gauge fields at late times were to provide the
initial conditions for a parton cascade~\cite{GeigerWang} or for
hydrodynamic evolution if it can be determined that the matter
produced has equilibrated~\cite{Bj83Josef97}.

The model is formulated in the infinite momentum frame $P^+\rightarrow
\infty$ and light cone gauge $A^+=0$. It contains a dimensionful parameter
$\mu^2$, defined to be the color charge squared per unit
area,
\begin{equation}
\mu^2 = {A^{1/3}\over {\pi r_0^2}} \int_{x_0}^1 dx \left({1\over 2
N_c} q(x,Q^2) + {N_c\over {N_c^2-1}} g(x,Q^2)\right) \, .
\end{equation}
Here $q,g$ stand for the {\it nucleon} quark and gluon structure
functions at the resolution scale $Q$ of the physical process of
interest. Also, above $x_0 = Q/\sqrt{s}$. Using the HERA structure
function data, Gyulassy and McLerran~\cite{MikLar} estimated that
$\mu\leq 1$ GeV for LHC energies and $\mu \leq 0.5$ GeV at RHIC. Thus
a window of applicability for weak coupling techniques does exist, and
higher order calculations will tell us if it is smaller or larger than
the simple classical estimate.

An interesting property of the light cone gauge is that final state
interactions are absent! Kovchegov and Mueller~\cite{KovMuell98}
showed that the effects of final state interactions, as seen in a
covariant gauge computation, are already contained in the nuclear
wavefunction in light cone gauge. This non--trivial observation is at
the heart of the approach described in this talk. Finally, we should alert 
the reader to alternative approaches to the one described here 
pursued in Refs.\cite{MullPoschl}--\cite{Makhlin}.

\section*{Classical model of gluon production}

At very high energies, the hard valence quark (and gluon) modes are highly
Lorentz contracted, static sources of color charge for the wee parton,
Weizs\"acker--Williams, modes in the nuclei. The sources are described
by the current
\begin{equation}
J^{\nu,a}(r_t) = \delta^{\nu +}\rho_{1}^a (r_t)\delta(x^-) + \delta^{\nu -}
\rho_{2}^a (r_t) \delta(x^+) \, ,
\label{sources}
\end{equation}
where $\rho_{1(2)}$ correspond to the color charge densities of the
hard modes in nucleus 1 (nucleus 2) respectively.  The classical field
describing the small x modes in the EFT is obtained by solving the Yang--Mills
equations in the presence of the two sources. We have then
\begin{equation}
D_\mu F^{\mu\nu} = J^\nu \, .
\label{yangmill}
\end{equation}
The small x glue distribution is simply related to the Fourier
transform $A_i^a (k_t)$ of the solution to the above equation by
$<A_i^a(k_t) A_i^a(k_t)>_\rho$.

The above averaging over the classical charge distributions is defined by
\begin{equation}
\langle O\rangle_\rho = \int d\rho_{1}d\rho_{2}\, O(\rho_1,\rho_2) 
 \exp\left( -\int d^2 r_t {{\rm Tr}\left[\rho_1^2(r_t)+\rho_2^2(r_t)
\right]
\over {2g^4\mu^2}}\right) \, .
\end{equation}
We have assumed identical nuclei with
equal Gaussian weights $g^4\mu^2$.

Before the nuclei collide ($t<0$), a solution of the equations of motion is
\begin{equation}
A^{\pm}= 0 \, ; \, A^i= 
\theta(x^-)\theta(-x^+)\alpha_1^i(r_t)+\theta(x^+)\theta(-x^-)
\alpha_2(r_t) \, ,
\label{befsoln}
\end{equation}
where $\alpha_{q}^i(r_t)$ ($q=1,2$ denote the labels of the nuclei and 
$i=1,2$ are the two transverse Lorentz indices)
are {\it pure gauge} fields defined through the gauge transformation
parameters $\Lambda_{q}(\eta,r_t)$~\cite{MikLar}
\begin{equation}
\alpha_{q}^i(r_t) = {1\over i}\left(Pe^{-i\int_{\pm \eta_{\rm proj}}^0
d\eta^\prime
\Lambda_{q}(\eta^\prime,r_t)}\right)
\nabla^i\left(Pe^{i\int_{\pm \eta_{\rm proj}}^0 d\eta^\prime 
\Lambda_{q}(\eta^\prime,r_t)}\right) \, .
\label{puresoln}
\end{equation}
Here $\eta=\pm\eta_{\rm proj}\mp\log(x^{\mp}/x_{\rm proj}^{\mp})$ is
the rapidity of the nucleus moving along the positive (negative) light
cone with the gluon field $\alpha_{1(2)}^i$. The
$\Lambda_{q}(\eta,r_t)$ are determined by the color charge
distributions $\Delta_\perp\Lambda_q=\rho_q$ (q=1,2) with
$\Delta_\perp$ being the Laplacian in the perpendicular plane.

For $t>0$ the solution is no longer pure gauge. Working in the Schwinger 
gauge $A^{\tau}\equiv x^+ A^- + x^- A^+ =0$, 
Kovner, McLerran  and Weigert~\cite{KLW} found that with the ansatz
\begin{equation}
A^{\pm}=\pm x^{\pm}\alpha(\tau,r_t)\,\,;\,\,
A^i=\alpha_\perp^i(\tau,r_t) \, ,
\label{ansatz}
\end{equation}
where $\tau=\sqrt{2x^+ x^-}$, Eq.~\ref{yangmill} could be written more 
simply in terms of $\alpha$ and $\alpha_\perp$. Note that these fields are
independent of $\eta$-the solutions are explicitly boost invariant in the 
forward light cone!

The initial conditions for the fields $\alpha(\tau,r_t)$ and
$\alpha_\perp^i$ at $\tau =0$ are obtained by matching the equations
of motion (Eq.~\ref{yangmill}) at the point $x^\pm =0$ and along the
boundaries $x^+=0,x^->0$ and $x^-=0,x^+>0$. Remarkably, for such
singular sources, there exist a set of non--singular initial
conditions for the smooth evolution of the classical fields in the
forward light cone. One obtains
\begin{equation}
\alpha_\perp^i|_{\tau=0}= \alpha_1^i+\alpha_2^i \,\,;\,\,
\alpha|_{\tau=0}={i\over 2} [\alpha_1^i,\alpha_2^i] \, .
\label{initial}
\end{equation}
Gyulassy and McLerran have shown~\cite{MikLar} that even when the fields
$\alpha_{1,2}^i$ before the collision are smeared out in rapidity, to 
properly account for singular contact terms in the equations of motion, the
above boundary conditions remain unchanged.
Further, since as mentioned the equations are very singular at $\tau=0$, the
only condition on the derivatives of the fields that would lead to regular
solutions are $\partial_\tau \alpha|_{\tau=0},\partial_\tau \alpha_\perp^i
|_{\tau=0} =0$.

Perturbative solutions of the Yang--Mills equations to order $\rho^2$
in the color charge density (or equivalently to second order in
$\alpha_S\mu/k_t$) were found, and at late times, after averaging over
the Gaussian sources, the number distribution of classical gluons was
found to be~\cite{KLW,MikLar,YuriDirk,SerBerDir}
\begin{equation} 
{dN\over {dyd^2 k_t}} = \pi R^2 {2g^6 \mu^4\over {(2\pi)^4}} {N_c (N_c^2-1)
\over k_t^4} L(k_t,\lambda) \, ,
\label{GunBer}
\end{equation}
where $L(k_t,\lambda)$ is an infrared divergent function at the scale
$\lambda$. This result agrees with the quantum bremsstrahlung formula
of Gunion and Bertsch~\cite{GunionBertsch}. Also, Guo has shown that the 
above result is equivalent to the perturbative QCD factorized result for 
the process $qq\rightarrow qqg$~\cite{Xiaofeng}.

From the above expression, it is clear that distributions are very
sensitive to $L(k_t,\lambda)$. Usually, as in Gunion and Bertsch, this
divergence is absorbed in a non--perturbative form factor. What is
novel about the classical approach is that, at sufficiently high
energies, the non--linearities in the Yang--Mills fields
self--consistently regulate this infrared divergence. To confirm this
claim, one needs to solve the Yang--Mills equations to all orders in
$\alpha_S \mu/ k_t$. A non--perturbative solution of the Yang--Mills
equations on a two dimensional transverse lattice was performed by 
us~\cite{alexraj} and is described below.

\section*{Real time simulations of Yang--Mills I: Lattice formulation}

We have seen above that the Yang--Mills equations are boost invariant. This 
is a consequence of the sources being $\delta$--functions on the light cone--
the nuclei are assumed to move with the speed of light! Since this is not 
the case, boost invariance is only approximate. It should, however, be a
good assumption at the energies of interest--especially at central rapidities.

The boost invariance assumption simplifies our numerical work
considerably. We now have a 2+1--dimensional theory and all the
dynamics is restricted to the transverse plane (we assume $\eta=0$).
The Yang--Mills equations are most conveniently solved by fixing
$A^\tau=0$ gauge and solving Hamilton's equations. Gauge invariance is
ensured by defining the theory, in the usual way, on a 2-dimensional
transverse lattice. The lattice Hamiltonian is the Kogut-Susskind
Hamiltonian for gauge fields coupled to an adjoint scalar
\begin{eqnarray}
H_L&=& {1\over{2\tau}}\sum_{l\equiv (j,\hat{n})} 
E_l^{a} E_l^{a} + \tau\sum_{\Box} \left(1-
{1\over 2}{\rm Tr} U_{\Box}\right)  \, ,\nonumber \\
&+& {1\over{4\tau}}\sum_{j,\hat{n}}{\rm Tr}\,
\left(\Phi_j-U_{j,\hat{n}}\Phi_{j+\hat{n}}
U_{j,\hat{n}}^\dagger\right)^2 +{\tau\over 4}\sum_j {\rm Tr}\,p_j^2,
\label{hl}
\end{eqnarray}
where $E_l$ are generators of right covariant derivatives on the group
and $U_{j,\hat{n}}$ is a component of the usual SU(2) matrices corresponding
to a link from the site $j$ in the direction $\hat{n}$. The first two terms
correspond to the contributions to the Hamiltonian from the chromoelectric and
chromomagnetic field strengths respectively. Also, above 
$\Phi\equiv \Phi^a\sigma^a$ is the adjoint scalar field with its conjugate
momentum $p\equiv p^a\sigma^a$.

Lattice equations of motion follow directly from $H_L$ of
Eq.~\ref{hl}.  For any dynamical variable $v$, with no explicit time
dependence, ${\dot v}=\{H_L,v\}$, where ${\dot v}$ is the derivative
with respect to $\tau$, and $\{\}$ denote Poisson brackets. We take
$E_l$, $U_l$, $p_j$, and $\Phi_j$ as independent dynamical variables,
whose only nonvanishing Poisson brackets are
$\{p_i^a,\Phi_j^b\}=\delta_{ij}\delta_{ab}$;
$\{E_l^a,U_m\}=-i\delta_{lm}U_l\sigma^a$;
$\{E_l^a,E_m^b\}=2\delta_{lm}\epsilon_{abc}E_l^c$ (no summing of
repeated indices). The equations of motion are consistent with a set
of local constraints (Gauss' laws).  Their evolution in $\tau$ after
the nuclear collision is determined by Hamilton's equations and their
values at the initial time $\tau =0$.

The initial conditions on the lattice are the constraints on the
longitudinal gauge potential $A^\pm$ and the transverse link matrices
$U_\perp$ at $\tau=0$.  The longitudinal gauge potentials are zero
outside the light cone and satisfy the Schwinger gauge condition
$A^\tau=0$ inside the light cone $x_\pm>0$. They can be written,
as in the continuum case (see Eq.~\ref{ansatz}), as
\begin{equation}
A^\pm=\pm x^\pm\theta(x^+)\theta(x^-)\alpha(\tau, x_t)\, .
\label{apm}
\end{equation}
The transverse link matrices are, for each nucleus, pure gauges before the
collision. This fact is reflected by writing
\begin{equation}U_\perp=\theta(-x^+)\theta(-x^-)I+\theta(x^+)\theta(x^-)U(\tau)
+\theta(-x^+)\theta(x^-)U_1+\theta(x^+)\theta(-x^-)U_2 \, ,
\label{uperp}
\end{equation}
where $U_{1,2}$ are pure gauge.

The pure gauges are defined on the lattice as follows. To each lattice
site $j$ we assign two SU($N_c$) matrices $V_{1,j}$ and
$V_{2,j}$. Each of these two defines a pure gauge lattice gauge
configuration with the link variables
$U_{j,\hat{n}}^q = V_{q,j}V_{q,j+n}^\dagger$
where $q=1,2$ labels the two nuclei. Also, as in the continuum, the gauge 
transformation matrices $V_{q,j}$ are determined by the color charge 
distribution $\rho_{q,j}$ of the nuclei, normally distributed with the 
standard deviation $g^4\mu_L^2$: 
\begin{equation}
P[\rho_q]\propto\exp\left(-{1\over{2g^4\mu_L^2}}\sum_j\rho_{q,j}^2\right).
\label{distriblat}
\end{equation}
Parametrizing $V_{q,j}$ as $\exp(i\Lambda^q_j)$ with Hermitean traceless
$\Lambda^q_j$, we then obtain $\Lambda^q_j$ by solving the lattice
Poisson equation
\begin{equation}
\Delta_L\Lambda^q_j\equiv\sum_n\left(\Lambda^q_{j+n}+\Lambda^q_{j-n}
-2\Lambda^q_j\right)=\rho_{q,j}.
\label{latpoi}
\end{equation}
The correct 
continuum solution (Eqs.~\ref{befsoln} and \ref{ansatz}) for the transverse 
fields $A_\perp$ is recovered by taking the formal continuum limit of 
Eq.~\ref{uperp}.

Using the general representation of the gauge fields in
Eqs.~\ref{apm} and~\ref{uperp}, we shall now state the initial conditions
for them at $\tau=0$. 
\begin{eqnarray}
U_\perp |_{\tau=0} &=& (U_1+U_2)(U_1^\dagger + U_2^\dagger)^{-1} \,\, ; 
\,\, E_l |_{\tau=0} = 0 \, . \nonumber \\
p_j |_{\tau=0} &=& 2\alpha \,\, ; \,\, \Phi_j = 0 \, ,
\label{dinitial}
\end{eqnarray}
where $U_\perp$ is defined as $\exp(ia_\perp\alpha_\perp)$.  The above
initial conditions are obtained by matching the lattice equations of
motion in the four light cone regions at $\tau =0$. For details we refer the 
reader to Ref.~\cite{alexraj}.

\section*{Real time simulations of Yang--Mills II: Discussion of results}

In Ref.~\cite{alexraj}, we reported results of simulations of the 
time evolution of classical fields in a 2+1--dimensional SU(2) 
gauge theory described by the Hamiltonian in Eq.~\ref{hl}.
The simulations were carried out on transverse lattices 
ranging from $20\times 20$ sites to $160\times 160$ sites. The lattice 
results depend on one dimensionless parameter, $g^2\mu L$~\cite{RajGavai}. 
The parameter $L$ corresponds to the size of the nucleus; $\mu$ defined in 
Eq.~(1) is determined by the size of the nucleus, the energy of the nucleus, 
and the hard scale $Q$ of interest; the strong coupling $g$ runs as a 
function of either $Q$ or $\mu$ depending on which is greater.

Thus, once given the energy and size of the incoming nuclei, and the
hard scale of interest, we can determine the evolution of gauge fields
in the central rapidity region. Consider, for instance, colliding two
gold nuclei (A$\sim$ 200) at RHIC and LHC energies. Approximating 
$L^2= \pi R^2$, we obtain $L=11.6$ fm. For the hard scales of interest,
$Q\approx$ 1--2 GeV, $\mu\sim 0.5$ GeV at RHIC ($\mu \sim 1$ GeV at
LHC). One then obtains
\begin{eqnarray}
g^2 \mu L \approx &120& \,\,\,\,{\rm (RHIC)} \nonumber \\ 
          \approx &240& \,\,\,\,{\rm (LHC)}  \, .
\label{coupling}
\end{eqnarray}
Above we have chosen $g=2$ (or equivalently, $\alpha_S\approx 0.3$).

On the lattice, the lattice coupling is $g^2 (\mu a) (L/a)\equiv g^2 \mu_L 
N$. The continuum limit is obtained by keeping $g^2\mu_L N$ fixed (to the 
physical value of interest-as in Eq.~\ref{coupling}) and taking $\mu_L$ to 
zero. It appears from our simulations that we are in the weak coupling 
regime for $\mu_L=0.017,0.035$ in lattice units. For the physical values of 
$g^2 \mu L$ above, these would correspond to lattices an order of magnitude 
larger than those considered so far. Detailed simulations on the above 
physical scenario will be reported at a later date.

We now turn to an issue of some concern; whether quantities of
interest have a continuum limit (in the above defined sense) in the
classical theory. For instance, in thermal field theories, it is not
clear that dynamical quantities such as auto--correlation functions
have a well defined limit as the lattice spacing $a\rightarrow 0$. In
the EFT described here, there is reason to be more optimistic.

Consider the following gauge invariant quantity; the 
energy density $p^a p^a = E_k/N^2$ of the scalar field on
the lattice (in units of $\mu_L^4$). It is plotted as a function of
the lattice size $N$ (in units of the lattice spacing) in Fig.~1 for 
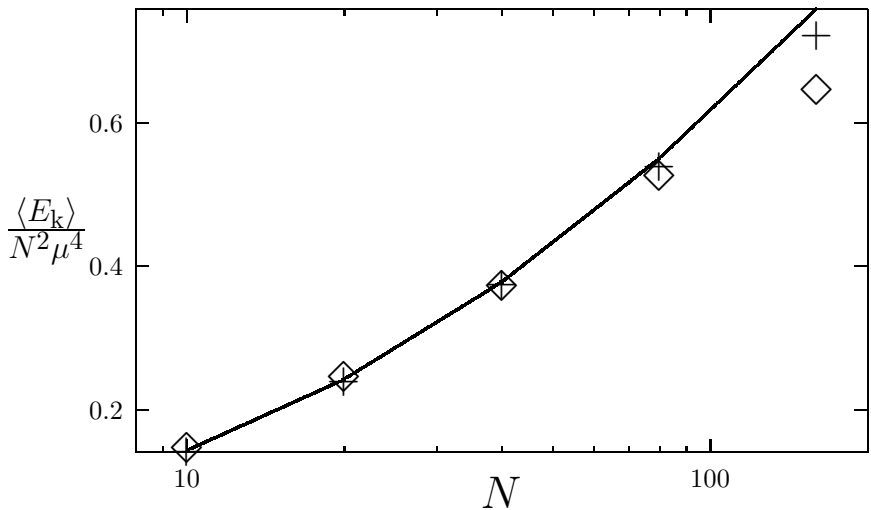
\begin{figure}[ht]
%\input  ekvsl
% GNUPLOT: LaTeX picture
\setlength{\unitlength}{0.240900pt}
\ifx\plotpoint\undefined\newsavebox{\plotpoint}\fi
\sbox{\plotpoint}{\rule[-0.200pt]{0.400pt}{0.400pt}}%
\begin{picture}(1350,900)(0,0)
\begin{Large}
\font\gnuplot=cmr10 at 10pt
\gnuplot
\sbox{\plotpoint}{\rule[-0.200pt]{0.400pt}{0.400pt}}%
\put(181.0,229.0){\rule[-0.200pt]{4.818pt}{0.400pt}}
\put(161,229){\makebox(0,0)[r]{0.2}}
\put(1310.0,229.0){\rule[-0.200pt]{4.818pt}{0.400pt}}
\put(181.0,455.0){\rule[-0.200pt]{4.818pt}{0.400pt}}
\put(161,455){\makebox(0,0)[r]{0.4}}
\put(1310.0,455.0){\rule[-0.200pt]{4.818pt}{0.400pt}}
\put(181.0,680.0){\rule[-0.200pt]{4.818pt}{0.400pt}}
\put(161,680){\makebox(0,0)[r]{0.6}}
\put(1310.0,680.0){\rule[-0.200pt]{4.818pt}{0.400pt}}
\put(181.0,163.0){\rule[-0.200pt]{0.400pt}{2.409pt}}
\put(181.0,849.0){\rule[-0.200pt]{0.400pt}{2.409pt}}
\put(223.0,163.0){\rule[-0.200pt]{0.400pt}{2.409pt}}
\put(223.0,849.0){\rule[-0.200pt]{0.400pt}{2.409pt}}
\put(261.0,163.0){\rule[-0.200pt]{0.400pt}{4.818pt}}
\put(261,122){\makebox(0,0){10}}
\put(261.0,839.0){\rule[-0.200pt]{0.400pt}{4.818pt}}
\put(508.0,163.0){\rule[-0.200pt]{0.400pt}{2.409pt}}
\put(508.0,849.0){\rule[-0.200pt]{0.400pt}{2.409pt}}
\put(653.0,163.0){\rule[-0.200pt]{0.400pt}{2.409pt}}
\put(653.0,849.0){\rule[-0.200pt]{0.400pt}{2.409pt}}
\put(756.0,163.0){\rule[-0.200pt]{0.400pt}{2.409pt}}
\put(756.0,849.0){\rule[-0.200pt]{0.400pt}{2.409pt}}
\put(835.0,163.0){\rule[-0.200pt]{0.400pt}{2.409pt}}
\put(835.0,849.0){\rule[-0.200pt]{0.400pt}{2.409pt}}
\put(900.0,163.0){\rule[-0.200pt]{0.400pt}{2.409pt}}
\put(900.0,849.0){\rule[-0.200pt]{0.400pt}{2.409pt}}
\put(955.0,163.0){\rule[-0.200pt]{0.400pt}{2.409pt}}
\put(955.0,849.0){\rule[-0.200pt]{0.400pt}{2.409pt}}
\put(1003.0,163.0){\rule[-0.200pt]{0.400pt}{2.409pt}}
\put(1003.0,849.0){\rule[-0.200pt]{0.400pt}{2.409pt}}
\put(1045.0,163.0){\rule[-0.200pt]{0.400pt}{2.409pt}}
\put(1045.0,849.0){\rule[-0.200pt]{0.400pt}{2.409pt}}
\put(1083.0,163.0){\rule[-0.200pt]{0.400pt}{4.818pt}}
\put(1083,122){\makebox(0,0){100}}
\put(1083.0,839.0){\rule[-0.200pt]{0.400pt}{4.818pt}}
\put(1330.0,163.0){\rule[-0.200pt]{0.400pt}{2.409pt}}
\put(1330.0,849.0){\rule[-0.200pt]{0.400pt}{2.409pt}}
\put(181.0,163.0){\rule[-0.200pt]{276.794pt}{0.400pt}}
\put(1330.0,163.0){\rule[-0.200pt]{0.400pt}{167.666pt}}
\put(181.0,859.0){\rule[-0.200pt]{276.794pt}{0.400pt}}
\put(41,511){\makebox(0,0){${{\langle E_{\rm k}\rangle}\over{N^2\mu^4}}$}}
\put(755,100){\makebox(0,0){$N$}}
\put(181.0,163.0){\rule[-0.200pt]{0.400pt}{167.666pt}}
\put(261,165){\raisebox{-.8pt}{\makebox(0,0){$\Diamond$}}}
\put(508,277){\raisebox{-.8pt}{\makebox(0,0){$\Diamond$}}}
\put(756,419){\raisebox{-.8pt}{\makebox(0,0){$\Diamond$}}}
\put(1003,592){\raisebox{-.8pt}{\makebox(0,0){$\Diamond$}}}
\put(1250,727){\raisebox{-.8pt}{\makebox(0,0){$\Diamond$}}}
\sbox{\plotpoint}{\rule[-0.400pt]{0.800pt}{0.800pt}}%
\put(261,165){\usebox{\plotpoint}}
\multiput(261.00,166.41)(1.106,0.501){217}{\rule{1.964pt}{0.121pt}}
\multiput(261.00,163.34)(242.923,112.000){2}{\rule{0.982pt}{0.800pt}}
\multiput(508.00,278.41)(0.811,0.501){299}{\rule{1.497pt}{0.121pt}}
\multiput(508.00,275.34)(244.893,153.000){2}{\rule{0.748pt}{0.800pt}}
\multiput(756.00,431.41)(0.637,0.500){381}{\rule{1.219pt}{0.121pt}}
\multiput(756.00,428.34)(244.471,194.000){2}{\rule{0.609pt}{0.800pt}}
\multiput(1003.00,625.41)(0.525,0.500){463}{\rule{1.041pt}{0.121pt}}
\multiput(1003.00,622.34)(244.840,235.000){2}{\rule{0.520pt}{0.800pt}}
\sbox{\plotpoint}{\rule[-0.500pt]{1.000pt}{1.000pt}}%
\sbox{\plotpoint}{\rule[-0.600pt]{1.200pt}{1.200pt}}%
\put(261,163){\makebox(0,0){$+$}}
\put(508,274){\makebox(0,0){$+$}}
\put(756,426){\makebox(0,0){$+$}}
\put(1003,612){\makebox(0,0){$+$}}
\put(1250,817){\makebox(0,0){$+$}}
\end{Large}
\end{picture}
\vskip-.6cm
\caption{The lattice size dependence of the scalar kinetic energy density,
expressed in units of $\mu^4$ for $\mu=0.0177$ (pluses) and $\mu=0.035$
(diamonds). The solid line is the LPTh prediction. The error bars are smaller
than the plotting symbols.}
\label{ekvsl}
\end{figure}
$\mu_L=0.0177, 0.035$. The solid line is the prediction from lattice
perturbation theory (LPTh.). It is given by 
\begin{equation}
p^a p^a = 6\left(\mu\over {N}\right)^4 \sum_{n,n^\prime} 
\left[ \left(\sum_{\vec{k}} {\sin(l_n)\sin(l_{n^\prime})\over \Delta^2 (l)}
\right)^2 + 16\left(\sum_{\vec{k}} {\sin^2({l_n\over 2})\sin^2({l_{n^\prime}
\over 2})\over \Delta^2 (l)}\right)^2 \right] \, ,
\label{dike}
\end{equation}
where $l_n = 2\pi n/N$ and $\Delta(l)= 2\sum_{n=1,2} (1-\cos(l_n))$ is the 
usual lattice Laplacian. The continuum limit of the above equation has the 
form $p^ap^a\longrightarrow A+B\log^2(L/a)$, where $A$ and $B$ are constants 
that can be determined from the above equation.

From Fig.~1, it can be seen that the kinetic energy density diverges
as $\log^2 (L/a)$--i.e., it does not have a well--defined continuum
limit. This divergence is softer than in a thermal theory, where the
energy density diverges as an inverse power of the lattice spacing. If
we subtract the perturbative result in Eq.~\ref{dike} at some scale
$\Lambda_{nonpert}$, the resulting expression will have a continuum
limit but will depend on $\Lambda_{nonpert}$. The trick is to find a
$\Lambda_{nonpert}$ for which the contribution to the \lq\lq
non--perturbative'' scalar kinetic energy density converges as 
$\mu_L\rightarrow 0$. In Table.~1, we 
show the results, at $\tau=0$, of simulations where $g^2\mu L = 33.6$ is 
kept fixed (note: here we fix $g=1$). The lattice cut--offs (recall 
that $k_t=2\pi n/L$) of $n_{cut} = 10,15,20$ correspond to $\Lambda_{nonpert}=
1.08,1.62,2.16$ GeV respectively.
\begin{table}[h]
\caption{Scalar k.e. density ${E_k\over{N^2\mu_L}}$ vs $\mu_L$ for  
fixed $g^2 \mu L = 33.6$. 
Columns 3--5 are the scalar.ke.d after subtracting LPTh. 
contribution for $n>n_{cut}$ in Eq.~\ref{dike}.}
\begin{tabular}{|c||c|c|c|c|}
\multicolumn{1}{|c|}{$\mu_L$} & \multicolumn{1}{c|}{${E_k\over {N^2\mu^4}}$} &
\multicolumn{1}{c|}{$n_{cut}$=10} & \multicolumn{1}{c|}{$n_{cut}$=15} & 
\multicolumn{1}{c|}{$n_{cut}$=20} \\ \hline
.84 &  07.24e-02 & .0341 & .0562 & .0702 \\ \hline
.56 &  10.20e-02 & .0407 & .0636 & .0791 \\ \hline
.42 &  12.38e-02 & .0426 & .0694 & .0852 \\ \hline
.21 &  17.60e-02 & .0268 & .0702 & .0945 \\ \hline
\end{tabular}
\end{table}
For $n_{cut}=15$, the non--perturbative contribution to the
scalar.ke.density appears to converge to a constant value as $\mu_L$
is decreased, keeping $g^2\mu L$ fixed. This result must of course be
confirmed for larger lattices and other values of $g^2\mu L$. Caveat
aside, our results appear to suggest that for a particular
$\Lambda_{nonpert}$ there is a non--perturbative contribution to the
energy density that survives in the continuum limit.

Presumably, the subtraction described above may also be performed for
static quantities in a thermal field theory. It is unlikely that this
procedure there is reliable for dynamic quantities. In our case, the scalar
kinetic energy is a dynamic quantity evolving in time. If the above
mentioned procedure is to be use, it must also be valid at later
times. We are optimistic that this may be the case because our
simulations suggest that the time evolution of hard modes decouples
from that of the soft modes. This is illustrated very clearly by
Fig.~2 of the most recent of our papers cited in Ref.~\cite{alexraj}.

The parameter $\Lambda_{nonpert}$ may be given a physical
interpretation. At sufficiently high energies, one may be able to
relate it to the physical scale at which one begins to see deviations
from perturbative predictions (note: at very high energies one expects
$\Lambda_{nonpert}>>\Lambda_{QCD}$) due to high parton density
effects.

We are currently studying whether the non--perturbative contribution
to the scalar energy density also has a robust continuum limit at late
times~\cite{alexraj2}.

\end{document}